\documentclass[prb,showpacs,showkeywords]{revtex4}
\usepackage{amsmath}
\usepackage{graphicx}

\input{tcilatex}

\begin{document}

\title{Magnetic phases of $t-J$ model on triangular lattice}
\author{Yongjin Jiang}
\affiliation{Center for Advanced Study, Tsinghua University, Beijing, 100084, China}
\author{Fan Yang}
\affiliation{Center for Advanced Study, Tsinghua University, Beijing, 100084, China}
\author{Tao Li}
\affiliation{Center for Advanced Study, Tsinghua University, Beijing, 100084, China}

\begin{abstract}
We study the magnetic properties of the $t-J$ model on triangular lattice in
light of the recently discovered superconductivity in Na$_{x}$CoO$_{2}$
system. \ We formulate the problem in the Schwinger Boson - slave Fermion
scheme and proposed a sound mean field ansatz(canting ansatz) for the RVB\
order parameters. Working with the canting ansatz, we map out the
temperature-doping phase diagram of the model for both sign of the hopping
term. We find the prediction of the $t-J$ model differ drastically from that
of earlier LSDA calculation and there is large doping range in which the
system show zero net magnetization, rather than saturated magnetization as
predicted in the LSDA calculation. We show the result of LSDA is unreliable
in the strong coupling regime due to its neglect of electron correlation. We
find the spin Berry phase play a vital role in this geometrically frustrated
system and the various states in the phases diagram are characterized(and
distinguished) by their respective spin Berry phase, rather than any
Landau-like order parameter related to broken symmetry. We find the spin
Berry phase is responsible for the qualitative difference in the low energy
excitation spectrum of the various states of the phase diagram. We argue the
phase boundary in the mean field phase diagram may serve as the first
explicit and realistic example for phase transition between states with
different quantum orders which in our case is nothing but the spin Berry
phase. We also find an exotic state with nonzero spin chirality but no spin
ordering is stable in a large temperature and doping range and find this
state support a nonzero staggered current loop in the bulk of the system. We
propose to study this exotic state in the hole doped Na$_{1-x}$TiO$_{2}$
system. As\bigskip\ a by-product of this study, we also achieve an
improvement over earlier Schwinger Boson mean field theory of the triangular
antiferromagnet by obtaining two gapless spinon modes. We find the small gap
of the third spinon mode is caused by quantum fluctuation and our theory
reproduce the linear spin wave theory(LSWT) result in the semiclassical
limit.
\end{abstract}

\maketitle

\section{Introduction}

The recent discovery of superconductivity below T$_{c}\simeq 4.5K$ in Na$%
_{x} $Co0$_{2}\cdot y$H$_{2}$O$(x\simeq 0.35,y\simeq 1.3)$ \cite{1}has
aroused much research interest. This material constitutes another example of
superconducting transition metal oxides with layered structure besides the
cuprates. The similarity between the two systems make it possible that the
study of the new superconductor may shed light on those difficult problems
in the study of the cuprates. In Na$_{x}$CoO$_{2}$ material, the Co$^{4+}$
ion possess a half spin and play a similar role as the Cu$^{2+}$ ion in
cuprates. However, the Co$^{4+}$ in Na$_{x}$CoO$_{2}$ form a triangular
lattice rather than square lattice as in the cuprates. From theoretical
point of view, triangular lattice system is even more interesting since it
is intrinsically frustrated.

Shortly after the discovery of superconductivity several theoretical
proposals are raised to address its mechanism\cite{2,3,4}. On the basis of
the mean field analysis on $t-J$ model, a $d+id\prime $ pairing state is
proposed by Barskran\cite{2}, while Tanaka et al proposed a triplet pairing
state based on symmetry consideration\cite{3}. Experimentally, the pairing
symmetry is still under debate\cite{5,6,7}. Besides superconductivity, this
system also exhibit nontrivial magnetic properties. In particular, a phase
transition at 22 $K$ from paramagnetic to weak ferromagnetic state is
reported in Na$_{0.75}$CoO$_{2}$\cite{8}. Experience in the study of
cuprates tell us that the magnetic degree of freedom may play an important
role in the low energy physics of this strongly correlated system.
Especially, it is important to find out the relation between the magnetic
properties and the superconductivity. Thus a comprehensive understanding of
the magnetic properties of this system is valuable. In this respect, Singh
calculated the magnetic properties of the system in the LSDA scheme\cite{9}
and found that the carrier in the conduction band is always fully polarized,
even in the half filled case. This result is inconsistent with the result of
the strong coupling analysis based on $t-J$ model. In the LSDA-like weak
coupling treatment, the tendency of the system toward ferromagnet state is
exaggerated due to the absence of electron correlation(the electron avoid
each other only through the Pauli exclusion principle which make the
ferromagnetic ordering of the electron spin energetically favorable). In
technical terms, the Stoner factor $1-UN(E_{F})$ overestimate the
ferromagnetic instability in the strongly correlated regime. In fact, as $U$
increase, the density of state at the Fermi energy is also depressed.
Especially, at the half filled case, $N(E_{F})$ may reduce to zero due to
the Mott insulator physics. Thus a through study of the magnetic properties
of the system in the strong coupling regime is deserved.

In this paper, we study the magnetic properties of the system within the$t-J$
model perspective. Complementary to the weak coupling picture, in the strong
coupling regime, it is the reduction of kinetic energy rather than the
potential energy that dominate the low energy physics and thus determine the
magnetic properties since the no double occupancy constraints in the $t-J$
model already take into account of the effect of the potential energy in the
zeroth order. Here the kinetic energy contains contribution from both real
charge transport of doped holes and the virtual hopping process that lead to
spin exchange. The real and virtual part of the kinetic energy are competing
with each other. The real kinetic energy favor ferromagnetic alignment of
the spin due to the Nagaoka physics while the virtual kinetic energy or the
superexchange favor antiferromagnetic alignment of spin since it is blocked
by Pauli exclusion principle. At half filling, the real process is depressed
and the virtual process dominate. Thus a fully polarized state is obviously
unfavorable in this case. In fact, the triangular $t-J$ model show zero net
magnetization at the half filling and is generally believed to be in a
coplanar 120$^{0}$ state with three sublattices\cite{10,11,12,13}(see Figure
1). How dose this state evolve with doping? This is the central problem we
want to address in this paper. We find the answer depend crucially on the
sign of the hoping term of the $t-J$ model(on a triangular lattice, the
particle hole symmetry is broken and the sign of the hopping term is
essential). A negative hopping matrix element will induce a $\pi $ -flux
around each triangular plaqutte of the lattice and frustrate the motion of
the charge carrier. The charge carrier will also see another flux due to the
spin Berry phase in a noncollinear spin background. The interplay between
the two result in a complex doping dependence of the magnetic properties
found in this work. Although the hopping term in the Na$_{x}$CoO$_{2}$
system is positive, we will consider both positive and negative sign case in
our paper for general interest.

In this paper, we treat the $t-J$ model in the slave Fermion - Schwinger
Boson formalism since it is convenient for the discussion of magnetic
properties. At half filling, the spin system is in a coplanar state with
three sublattices. The spin in the three sublattices lay symmetrically on
their common plane with a 120$^{0}$ angle between each other. From this
state, the most natural way for the system to obtain a net magnetization is
to tilt the spin of the three sublattices symmetrically above their common
plane. This is supported by a unrestricted search of the semiclassical
energy functional. The state so obtained is called a canting state. With
this picture in mind, we propose a sound ansatz for the mean field RVB order
parameter for the spins. We call the mean field ansatz constructed in this
way the canting ansatz. Working with the canting ansatz, we map out the
whole doping-temperature phase diagram of the $t-J$ model for both sign of
the hopping term. The main result result of this paper can be summarized as
follow.

We find there is large doping range in which the system show zero net
magnetization, rather than saturated magnetization as predicted in the LSDA
calculation. Thus the LSDA result is unreliable in the strong coupling
regime.

We find the spin Berry phase play a vital role in this geometrically
frustrated system. We find the various states in the phase diagram are
characterized(and distinguished) by their respective spin Berry phase,
rather than any Landau-like order parameter related to broken symmetry. We
also find the spin Berry phase is responsible for the qualitative difference
in the low energy excitation spectrum of the various states in the phase
diagram. We argue that the phase boundary in the mean field phase diagram
are true phase transition lines and may serve as the first explicit and
realistic example for phase transition between states with different quantum
orders\cite{14}. We argue the spin Berry phase provide a first realistic
example for the concept of quantum order.

We find there exists an exotic state with nonzero spin chirality but no spin
ordering in a large temperature and doping range. We find this exotic state
support nonzero staggered current loop in the bulk of the system and propose
to study this exotic state in the hole doped Na$_{1-x}$TiO$_{2}$ system.

As\bigskip\ a by-product of this study, we also achieve an improvement over
earlier Schwinger Boson mean field theory on the triangular antiferromagnet.
The canting ansatz proposed in this paper lead to two branches of gapless
spinon modes. We find the small gap of the third spinon mode is caused by
quantum fluctuation and our theory reproduce the linear spin wave theory
result in the semiclassical limit.

Finally, we find our general phase diagram is consistent with experimental
result on Na$_{1-x}$CoO$_{2}$ system and a singlet pairing picture may be
more relevant in the small doping regime.

The paper is organized as follows. In section II, the Schwinger Boson-slave
Fermion representation for the $t-J$ model and our working ansatz, the
canting ansatz are introduced. The phase diagram is presented in section
III. In section IV, we present the result for the doping dependence of
magnetization and excitation spectrum. In the concluding section IV,
relevance of our result to experiments are discussed.

\bigskip

\section{ The Canting Ansatz}

\bigskip Our starting point is the $t-J$ model on triangular lattice,

\begin{equation}
H=-t\sum_{<i,j>,\sigma }(\hat{C}_{i\sigma }^{\dagger }\hat{C}_{j\sigma
}+H.c.)+J\sum_{<i,j>}\hat{S}_{i}\hat{S}_{j}  \label{1}
\end{equation}%
in which $\hat{C}_{i\sigma }$ satisfy the no double occupation constraint $%
\sum_{\sigma }\hat{C}_{i\sigma }^{+}\hat{C}_{i\sigma }\leq 1$. $%
\dsum\limits_{\left\langle i,j\right\rangle }$ denotes sum over
nearest-neighbouring sites. For the triangular lattice, the sign the $t$ is
essential and we will consider both sign of $t$ for theoretical interest,
although $t$ is positive in Na$_{x}$CoO$_{2}$ system.

At half filling, the model reduce to the much studied antiferromagnetic
Heisenberg model on a triangular lattice. It is believed that this model
posses a coplanar magnetic order with three sublattices at zero temperature%
\cite{10,11,12,13}. In this state, the spins in the three sublattices reside
symmetrically on their common plane with 120$^{0}$ angle between each
other(see Figure 1). This result can be obtained in the Schwinger-Boson mean
field theory of the model\cite{12,13}. In this formalism, the spin operator $%
\hat{S}_{i}$ is represented by Schwinger Boson $b_{i,\alpha }$

\begin{equation*}
\hat{S}_{i}=\frac{1}{2}b_{i\alpha }^{\dagger }\sigma _{\alpha \beta
}b_{i\beta }
\end{equation*}

with the constraint $\sum_{\alpha }b_{i\alpha }^{\dagger }b_{i\alpha }=1$
and the Hamiltonian is

\begin{equation*}
H_{s}=\frac{1}{4}J\sum_{<i,j>}b_{i\alpha }^{\dagger }\sigma _{\alpha \beta
}b_{i\beta }b_{j\gamma }^{\dagger }\sigma _{\gamma \delta }b_{j\delta }
\end{equation*}%
In the mean field treatment, order parameter $D_{ij}=\left\langle
b_{i\uparrow }^{{}}b_{j\downarrow }^{{}}-b_{i\downarrow }^{{}}b_{j\uparrow
}^{{}}\right\rangle $, $Q_{ij}=\left\langle \dsum\limits_{\alpha }b_{i\alpha
}^{\dagger }b_{j\alpha }^{{}}\right\rangle $ are introduced to represent the
spin correlation in the system. Although $D_{ij}$ and $Q_{ij}$ are
themselves spin rotational invariant, by condensing the bosonic spinon $%
b_{i\alpha }$ we can still break this symmetry and describe magnetic ordered
state(note both $D_{ij}$ and $Q_{ij}$ are needed for a nonbipartite lattice
to find the absolute minima of the energy functional). The choice of the
mean field ansatz $D_{ij}$ and $Q_{ij}$ depends on our understanding of the
ground state. The ansatz used in previous theories\cite{11,12,13} on this
problem can reproduce the three sublattice magnetic order, but fails to give
three branches of gapless spin wave and are thus not fully satisfactory. In
the following, we will determine the mean field ansatz with a semiclassical
analysis. When the system is doped, we should introduce another slave
particle to represent the hole degree of freedom. In this paper, we use the
slave Fermion- Schwinger Boson representation. In this formalism

\begin{equation*}
\hat{C}_{i\sigma }^{\dagger }=f_{i}b_{i\sigma }^{\dagger },
\end{equation*}%
in which $f_{i}$ is Fermionic operator for the holon, and $b_{i\sigma
}^{\dagger }$ is bosonic operator for the spinon just defined. The no double
occupation constraint is now expressed as: 
\begin{equation*}
f_{i}^{\dagger }f_{i}+\sum_{\sigma }b_{i\sigma }^{\dagger }b_{i\sigma }=1,
\end{equation*}%
and the Hamiltonian (\ref{1}) is: 
\begin{eqnarray}
H &=&-t\sum_{<i,j>}(f_{i}f_{j}^{\dagger }b_{i\sigma }^{\dagger }b_{j\sigma
}+h.c.)+\frac{1}{4}J\sum_{<i,j>}b_{i\alpha }^{\dagger }\sigma _{\alpha \beta
}b_{i\beta }b_{j\gamma }^{\dagger }\sigma _{\gamma \delta }b_{j\delta } 
\notag \\
&&+\sum_{i}\lambda _{i}(f_{i}^{\dagger }f_{i}+b_{i\sigma }^{\dagger
}b_{i\sigma }-1)-\mu \sum_{i}f_{i}^{\dagger }f_{i},  \label{3}
\end{eqnarray}

in which $\lambda _{i}$ is the Lagrangian multipliers to enforce particle
number constraint and $\mu $ is the chemical potential for the hole. This
Hamiltonian can be decoupled into bilinear form with the introduction of the
mean field order parameter $D_{ij}$ and $Q_{ij}$. In principle, we should
also include the order parameter $F_{ij}=\left\langle f_{i}^{\dagger
}f_{j}\right\rangle $ in the doped case. However, since the hole has no
dynamics of its own, $F_{ij}$ is proportional to $Q_{ij}^{{}}$ and is not an
independent variable. After the decoupling, we get the mean field Hamiltonian

\begin{equation*}
H_{s}=\dsum\limits_{\left\langle i,j\right\rangle }\left[ (tF_{ij}^{\ast }+%
\frac{J}{4}Q_{ij}^{\ast })b_{i\alpha }^{\dagger }b_{j\alpha }^{{}}-\frac{J}{4%
}D_{ij}^{\ast }\epsilon _{\alpha \beta }b_{i\alpha }^{\dagger }b_{j\beta
}^{\dagger }+H.c.\right] -\mu _{b}\dsum\limits_{i}b_{i\alpha }^{\dagger
}b_{i\alpha }^{{}}
\end{equation*}%
for spinon and

\qquad 
\begin{equation*}
H_{h}=\dsum\limits_{\left\langle i,j\right\rangle }\left[
tQ_{ij}^{{}}f_{j}^{\dagger }f_{i}+H.c.\right] -\mu
_{f}\dsum\limits_{i}f_{i}^{\dagger }f_{i}
\end{equation*}%
for holon. Here $\epsilon _{\alpha \beta }$ is the antisymmetric tensor. $%
\mu _{b}$ and $\mu _{f}$ are the spinon and holon chemical potential to
enforce the particle number constraints $\left\langle b_{i\alpha }^{\dagger
}b_{i\alpha }^{{}}\right\rangle =1-x$ and $\left\langle f_{i}^{\dagger
}f_{i}\right\rangle =x$ ,where $x$ is the hole concentration. From this
Hamiltonian, we see the spin background affect the hole motion through the
order parameter $Q_{ij}$ whose phase is nothing but the spin Berry phase
mentioned in the Introduction.

Now we determine the mean field ansatz $D_{ij}$ and $Q_{ij}$ from a
semiclassical analysis\cite{15}. In the semiclassical limit, we can take $%
b_{i\alpha }$ as a two component spinor

\begin{equation}
b_{i}=\sqrt{S_{{}}}\left( 
\begin{array}{c}
e^{-i\varphi _{i}}\cos (\frac{1}{2}\theta _{i}) \\ 
\sin (\frac{1}{2}\theta _{i})%
\end{array}%
\right)
\end{equation}

in which $\theta _{i}$ and $\varphi _{i}$ are spherical coordinates of the
spin $S_{i}$. Here an arbitrary gauge phase factor $e^{i\chi _{i}}$ is
omitted. Now let us consider possible magnetization of the system from the
coplanar state at half filling. In this parent state, the spins in the three
sublattices lay symmetrically in their common plane. Hence the most natural
way for the system to obtain a net magnetization is to tilt the spins of the
three sublattices symmetrically above their common plane. We call such a
state a canting state. In this state, the semiclassical spin on the three
sublattices are%
\begin{eqnarray}
b_{{}}^{A} &=&\sqrt{S}\left( 
\begin{array}{c}
\cos (\frac{1}{2}\theta ) \\ 
\sin (\frac{1}{2}\theta )%
\end{array}%
\right)  \notag \\
b_{{}}^{B} &=&\sqrt{S}\left( 
\begin{array}{c}
e^{-i\frac{2\pi }{3}}\cos (\frac{1}{2}\theta ) \\ 
\sin (\frac{1}{2}\theta )%
\end{array}%
\right)  \label{6} \\
b_{{}}^{C} &=&\sqrt{S}\left( 
\begin{array}{c}
e^{-i\frac{4\pi }{3}}\cos (\frac{1}{2}\theta ) \\ 
\sin (\frac{1}{2}\theta )%
\end{array}%
\right) ,  \notag
\end{eqnarray}%
where $A,B,C$ stands for the three sublattices. The coplanar state at half
filling is a special case of the canting state with $\theta =\frac{\pi }{2}$%
. This canting ansatz is supported by minimizing the semiclassical energy
functional. This energy functional is obtained by substituting (\ref{3})
into (\ref{2})(with $\sqrt{S}$ replaced by $\sqrt{S(1-\delta )}$ in the
doped case, where $\delta $ is the hole concentration) and then diagonalize
the holon Hamiltonian. We performed an unconstraint search on a $12\times 12$
lattice and find the canting state always has the lowest energy. In the
semiclassical limit, the mean field order parameter can be readily written
down. For the canting state, we have

\begin{eqnarray}
D_{AB} &=&\left\langle b_{A\uparrow }^{{}}b_{B\downarrow
}^{{}}-b_{A\downarrow }^{{}}b_{B\uparrow }^{{}}\right\rangle =D_{BC}e^{i%
\frac{4}{3}\pi }=D_{CA}e^{-i\frac{4}{3}\pi }=D  \label{4} \\
Q_{AB} &=&\left\langle \dsum\limits_{\alpha }b_{B\alpha }^{\dagger
}b_{A\alpha }^{{}}\right\rangle =Q_{BC}=Q_{CA}=Q  \notag
\end{eqnarray}%
To fix the gauge totally, we require $D$ to be real. In this paper, we will
study mean field ansatz satisfying (\ref{4}) and call such a mean field
ansatz canting ansatz. In this ansatz, a real order parameter $D$ and a
complex order parameter $Q$ are introduced to represent the spin correlation
in the system. In particular, the phase of $Q$ equals to one sixth of the
solid angle spanned by the spins in the three sublattices and is thus
related to the spin chirality and the spin Berry phase of the background
spins\cite{16}.

The mean field Hamiltonian with the canting ansatz can be diagonalized.
Introducing $\Psi _{k}=%
\begin{array}{ccc}
(f_{k}^{A} & f_{k}^{B} & f_{k}^{C})%
\end{array}%
^{T}$ and $\Phi _{k}=%
\begin{array}{cccccc}
(b_{k\uparrow }^{A} & b_{k\uparrow }^{B} & b_{k\uparrow }^{C} & 
b_{-k\downarrow }^{A\dagger } & b_{-k\downarrow }^{B\dagger } & 
b_{-k\downarrow }^{C\dagger })%
\end{array}%
^{T}$, the Hamiltonian can be written as

\begin{eqnarray*}
H_{s} &=&\sum_{k}\Phi _{k}^{\dagger }\mathbf{M}_{b}\Phi _{k} \\
H_{h} &=&\sum_{k}\Psi _{k}^{\dagger }\mathbf{M}_{h}\Psi _{k}
\end{eqnarray*}%
where the expression for the $6\times 6$ matrix $\mathbf{M}_{b}$ and the $%
3\times 3$ matrix $\mathbf{M}_{h}$ are given in the Appendix. Here the
momentum sum runs over the reduced Brillouin zone for the three sublattice
system. The six branches of spinon dispersions are

\begin{equation*}
\omega _{n}^{\pm }(k)=\left\vert -2\func{Im}(T)\func{Im}(\Gamma
_{n,k}^{{}})\pm \sqrt{\left[ \mu _{b}+2\func{Re}(T)\func{Re}(\Gamma
_{n,k}^{{}})\right] ^{2}-\frac{1}{4}\left[ D\func{Im}(\Gamma _{n,k}^{{}})%
\right] ^{2}}\right\vert
\end{equation*}

with $n=0,1,2$. The three branches of holon dispersions are

\begin{equation*}
\varepsilon _{n}^{{}}(k)=\mu _{f}+t\func{Re}(Q\Gamma _{n,k})
\end{equation*}

with $n=0,1,2$. Here $T=(-tF+\frac{Q}{4})e^{i\frac{2\pi }{3}}$ , $\Gamma
_{n,k}^{{}}=\gamma _{k}e_{{}}^{i\frac{2n\pi }{3}}$ and $\gamma
_{k}=\dsum\limits_{\delta }e^{ik\delta }$ in which $\delta =(1,0),(-\frac{1}{%
2},\pm \frac{\sqrt{3}}{2}).$ Since $\omega _{n}^{+}(k)=\omega _{n}^{-}(-k)$,
each branch of spinon mode is two-fold degenerate. This is a reflection of
the spin rotational symmetry which is unbroken in our treatment.

By solving the self-consistent mean field equations, we can obtain a series
of solutions for $D$ and $Q$. These solutions are characterized by the phase
of $Q$ . As we have shown, this phase angle, $\varphi $, equals to one sixth
of the solid angle spanned by the spins. Since the solid angle is defined
modulo $4\pi $ and the problem is symmetric under the reflection about the
common plane of the spins, we can restrict ourselves to the interval $0\leq
\varphi \leq \frac{\pi }{3}$. In the following, the state with $\varphi =0$
will be called a collinear state and the state with $\varphi =\frac{\pi }{3}$
will be called the coplanar state. A state with $\varphi $ between this two
limits will be called the canted state. In the next section, we will map out
the mean field phase diagram of the model by determining the solution with
the lowest free energy.

\bigskip

\section{The phase diagram}

\subsection{The $t>0$ case}

\bigskip The temperature-doping phase diagram for $t/J=1.5$ is shown in
Figure 2. At low doping, the coplanar state has the lowest free energy.
Above a critical doping about $0.5$ the system transform into the collinear
state through a first order phase transition. The canted state is never
stable for $t>0$. This result can be understood as follows. From (\ref{1}),\
we see the bare hopping term of the hole (which is $-t$) is renormalized by
the spin dependent factor $Q$ . For $t>0$, the hole will feel a flux of
strength $\pi \pm 3\varphi $ around each elementary triangular loop of the
lattice($\varphi $ is the phase of $Q$) and is in general frustrated. When $%
\varphi =\frac{\pi }{3}$, or, when the spin is in the coplanar state, the
frustration due to the bare hopping term is released by the spin Berry phase
from $Q$. Thus at low doping, the spins prefer to stay in the coplanar
state(this state is of course also favored by the spin exchange energy). At
high doping, the situation is reversed and now it is the 'holes'\bigskip\ of
the nearly filled hole band that are moving. The sign of the hopping term of
these 'holes' is opposite to that of the original hole and thus the
frustration due the bare hopping term is absent. Thus at high doping, the
spins prefer to align themselves ferromagnetically(the spin exchange energy
is suppressed at high doping due to dilution of spins). In the argument
raised above, we neglect the effect of the amplitude of the $Q$.
Semiclassical analysis shows that this neglect has no qualitative effect.

At zero temperature, the condensation of the spinon will lead to long range
order of spins. In the coplanar state, there are two gapless spinon
modes(see discussion in section IV). Both modes condense into the coplanar
spin long range order. In the collinear state, there is only one gapless
spinon mode. The condensation of this mode give rise to ferromagnetic long
range order of the spins. Thus unlike the LSDA result, there is a finite
doping range in our phase diagram in which the net magnetization is zero.

At finite temperature, the spin long range order does not exist. The
distinction between the coplanar state and the collinear state is now more
subtle. On symmetry ground, it is hard to tell apart between the two. In our
theory, the two states are distinguished by their spin Berry phase. The
problem here is whether the phase boundary between the coplanar state and
the collinear state represent true phase transition or just crossover. We
think this phase boundary still represent true phase transition since there
is a discontinues change of the spin correlation pattern(embodied in the
spin Berry phase) across it. This discontinues change in the spin
correlation pattern lead to a discontinues change in the excitation spectrum
and thus thermodynamical properties of the system across the phase boundary.
As will be shown in the next section, the spin Berry phase is responsible
for the qualitative difference in the excitation spectrum of the various
states in the phase diagram. In particular, in the coplanar state with $%
\varphi =\frac{\pi }{3}$, there is always two branches of gapless spinon(the
gap due to finite temperature is exponentially small), both of which have
linear dispersion. While in the collinear state (or any states with $\varphi
\neq \frac{\pi }{3}$), there is only one branch of gapless spinon whose
dispersion is quadratic. In other words, the two branches of gapless spinon
mode in the coplanar state seems to be protected by some kind of order not
related to any broken symmetry. This special kind of the order, which is
nothing but the characteristic spin Berry phase of the coplanar state, is
broken across the phase boundary between the coplanar state and the
collinear state. It is in this sense that we think the phase boundary a true
phase transition line. The concept of an order with no broken symmetry is
first advocated by Wen who give it the name quantum order\cite{14}. The
quantum order, as it defined by Wen, describes the structure in the quantum
wave function which is beyond the classification on symmetry ground and can
be detected by checking the excitation spectrum of the system which is
sensitive to the structure of the ground state wave function. In our case,
the quantum order is nothing but the spin Berry phase. We think the
transition between the coplanar state and the collinear state may serve as
the first explicit and realistic example for phase transition between states
with different quantum orders. This is the most important finding of this
paper. In this regard, it is seems natural to look at the zero temperature
transition between the two states also as a transition between different
quantum orders rather than the conventional Landau-type phase transition,
although the symmetry is really broken in this case.

\bigskip The realistic value of $t/J$ for Na$_{x}$CoO$_{2}$ is still
unsettled. In our theory, the phase boundary between the coplanar state and
the fully polarized state will shift to lower doping with increasing $t/J$.
If our assignment of $t/J=1.5$ is realistic(ARPES experiment report a
similar value\cite{17}), then the superconductivity observed around $x\sim
0.35$ is covered totally in the coplanar state in which ferromagnetic
fluctuation is suppressed. This implies that the singlet pairing picture may
be more relevant for the superconductivity in this system. Furthermore, the
ferromagnetic transition observed at $22K$ for $x\sim 0.75$ \cite{8}seems to
be also consistent with our phase diagram.\ 

\bigskip 

\subsection{\textbf{The }$t<0$ \textbf{case }}

The temperature-doping phase diagram for $t/J=-1.5$ is shown in Fig.3. The
phase diagram is much more complex as compared with that of the $t>0$ case.
At zero temperature, the spins cant gradually out of their common plane with
increasing doping until a critical doping level, above which the spin return
back into the coplanar state through a first order phase transition. At
finite temperature, the phase diagram show complex structures. Below a lower
critical doping, the zero temperature canted state return back into the
coplanar state through a second order phase transition at high temperature.
Above the lower critical doping, the canted state transform into the
collinear state through a second order phase transition. The coplanar state
at high doping also transform into the collinear state at high temperature,
but through a first order phase transition. The phase diagram contains both
first order and second order phase transition lines with tricritical and
quadricritical points join them.

This complex phase diagram structure can also be understood on semiclassical
ground. For $t<0$, the frustration due to the bare hopping term is absent
and the kinetic energy of the hole favor a state with $\varphi =0$, or, the
collinear state at low doping. The antiferromagnetic exchange energy act
against this tendency. The compromise between the two result in the canted
state in which the spin cant gradually out of the coplanar state at half
filling. At high doping, the situation is again reversed and now the kinetic
energy favors the coplanar state which is also favored by the exchange
energy. Below the lower critical doping, the spin canting at zero
temperature induced by the hole will eventually be erased by the thermal
fluctuation\ and the spins will return to the coplanar state at high
temperature. At higher doping level, the hole system is more robust and it
in fact help to enhance the spin correlation against thermal fluctuation.
This is quite different from the situation in cuprates. In the triangular
lattice, the spin is already frustrated at half filling. Doping a small
concentration of holes in fact help to release such geometrical frustration.
This special effect will be discussed further in the next section.

At finite temperature, the spin long range order can not exist and the
various phases in the phase diagram are characterized by the value of the
spin Berry phase. Among these phases, of particular interest is the canted
phase which hold a special kind of long range order even at finite
temperature. In the canted phase, the time reversal symmetry is
spontaneously broken and system posses a nonzero spin chirality. Thus the
phase boundary between the canted phase and other phases are true phase
transitions rather than crossovers even in the Landau sense. In the canted
phase, the spin chirality will induce a nonzero current of the hole. Since
the spin chirality is staggered, the induced current is also staggered. Thus
in the canted state, we expect staggered current to flow around each
triangular loop. This is the most important characteristic of the canted
phase. A detection of such current loop is interesting.

Before closing this subsection we note that although the $t<0$ case of the
model is not directly relevant to the Na$_{1-x}$CoO$_{2}$ system, it is
proposed that it may be appropriate for the hole doped system Na$_{1-x}$TiO$%
_{2}$\cite{4}.

\section{Physical Observable.}

\subsection{The doping dependence of the net magnetization.}

At zero temperature, the gapless spinon will condense into magnetic long
range order. For the canted state and the collinear state, this condensation
will lead to a nonzero net magnetization of the system. For the $t>0$
case(which is relevant for the Na$_{1-x}$CoO$_{2}$ system), there is
discontinues jump of the magnetization at the transition point between the
coplanar state and the collinear state. For $t<0$, the magnetization build
up gradually with increasing doping as the spin cant out of their common
plane and jump to zero when the spin return back into the coplanar state at
high doping. These results are shown in Figure 4. We also presents the
result for the fraction of the condensed spin (which is a measure of
magnitude of sublattice magnetization)in this figure. We see doping a small
amount of hole enhance the sublattice magnetization for both $t>0$ and $t<0$%
. The reason for this enhancement is that the doped hole can release
partially the geometric frustration inherent of the triangular lattice.

\bigskip 

\subsection{\protect\bigskip The spinon excitation spectrum}

In quantum systems, the structures in the ground state wave function often
have nontrivial consequences in the excitation spectrum of the system. In
fact, the excitation spectrum serve as the most direct way to detect any
non-symmetry-breaking related structure(or quantum order) in the ground
state wave function of quantum system. In our case, the various phases in
the phase diagram are distinguished by their spin Berry phase. Now we
discuss the effect of the spin Berry phase on the excitation spectrum of the
system.

In the reduced Brillouin Zone of the three sublattice system, there are
three branches of nondegenerate spinon modes. These three branches of
nondegenerate spinon modes transform into each other under the translation
in the momentum space by a reciprocal lattice vector of the reduced
Brillouin zone(this can be easily shown since $\gamma _{k+G}=\gamma
_{k}e^{\pm i\frac{2\pi }{3}}$, where $G$ is a reciprocal lattice vector of
reduced Brillouin zone). Thus in the extended zone scheme, these three
branches of spinon modes can be expressed by a single formula

\begin{equation}
\omega (k)=\left\vert 2\func{Im}(T)\func{Im}(\gamma _{k})+\sqrt{\left[ \mu
_{b}+2\func{Re}(T)\func{Re}(\gamma _{k})\right] ^{2}-\frac{1}{4}\left[ D%
\func{Im}(\gamma _{k})\right] ^{2}}\right\vert  \label{5}
\end{equation}

In the coplanar state $T$ is real(since $\varphi =\frac{\pi }{3}$) and the
spinon spectrum can be further reduced to

\begin{equation*}
\omega (k)=\sqrt{\left[ \mu _{b}+2T\func{Re}(\gamma _{k})\right] ^{2}-\frac{1%
}{4}\left[ D\func{Im}(\gamma _{k})\right] ^{2}}
\end{equation*}%
\qquad\ This spectrum is degenerate at the six corners of the Brillouin zone
of the triangular lattice($\gamma _{k}=3e^{\pm i\frac{2\pi }{3}}$at these
points). This degeneracy is the most important characteristic of the
excitation spectrum of the coplanar state and it is protected by the spin
Berry in this state. When $\varphi $ deviate from $\frac{\pi }{3}$, this
degeneracy is gone. At zero temperature, the spinon at these momentums will
become gapless and condense into the $120^{0}$ spin long range order. Thus,
at general doping there are two branches of gapless spin wave excitation on
the ordered spin background(the six corners of the Brillouin zone correspond
to two independent momentums). Since $\gamma _{k}\sim 3+uk^{2}$ for small $k$%
, both of the two modes have linear dispersion around the gapless point.
Besides these two gapless modes, there is a third minimum of the excitation
spectrum at the center of the Brillouin zone. The energy of this mode
decrease linearly with decreasing doping but remain finite at half filling
as shown Figure 5. The excitation spectrum of the half filled system is
shown in Figure 6 and it is compared with spin wave spectrum obtained in the
linear spin wave theory(LSWT)\cite{10}. The LSWT predict three branches of
gapless spin wave on the three sublattice spin background while our
Schwinger Boson mean theory predict only two. The reason for this
discrepancy can be seen in the following way. Requiring both the zone corner
modes and the zone center to be gapless, we obtain

\begin{equation*}
D^{2}=3\left\vert Q\right\vert ^{2}
\end{equation*}%
at half filling, a relation valid for coplanar state in the semiclassical
limit($S\longrightarrow \infty $ limit). Thus the zone center mode will also
become gapless in the semiclassical limit and our theory can reproduce the
result of the LSWT in this limit. This analysis also indicate that the gap
of the zone center mode is due to quantum fluctuation and may be observable
for the spin $\frac{1}{2}$ system. We note previous Schwinger Boson mean
field theory on the triangular antiferromagnet(with different mean field
ansatz)predict only one gapless mode\cite{11,12,13}. Thus our result is an
improvement over these earlier ones. At finite doping, the gap due to
quantum fluctuation is masked by the much larger gap due to the hole
background. However, the gaplessness at the zone corner is still intact as
long as the system remain in the coplanar state. This provide a explicit
example of protected excitation spectrum due to quantum order. Here the
quantum order is nothing but the spin Berry phase.

In the canted state and the collinear state(the collinear state is a special
case of the canted state), the degeneracy between the zone corner modes is
broken. From (\ref{5}), we see the splitting between the zone corner modes
is $\sqrt{3}\func{Im}(T)$ which increase with increasing canting angle.
Especially, in the collinear state in which $D=0,\varphi =0$, one of the
zone corner mode become degenerate with the massive zone center mode. For
general canting angle there are three nondegenerate modes in which only one
(at zone corner)become gapless at zero temperature. Since $\func{Im}(T)\func{%
Im}(\gamma _{k})\neq 0$ at the gapless point, the dispersion around it is
quadratic. The quadratic dispersion is a characteristic of the ferromagnet
spin wave. In our case, canting induce a weak ferromagnetic moment. The
gaplessness of the quadratic mode indicate that the long wavelength
fluctuation of the weak ferromagnetic moment is unaffected by the hole
background. The excitation spectrum for a general canting angle is shown in
Figure 7.

\section{\protect\bigskip Conclusion}

Now let's summarize the results of this paper. In this paper, we have
studied the magnetic properties of the $t-J$ model on triangular lattice in
light of the recently discovered superconductivity in Na$_{x}$CoO$_{2}$
system. \ We formulated the problem in the Schwinger Boson - slave Fermion
scheme and proposed a sound mean field ansatz, namely the canting ansatz,
for the RVB\ order parameters of spins. With the canting ansatz, we have
mapped out the temperature-doping phase diagram of the model for both sign
of the hopping term. We find the prediction of the $t-J$ model differs
drastically from that of LSDA calculation. In our theory, there is large
doping range in which the system show zero net magnetization, rather than
saturated magnetization as predicted in the LSDA calculation. We show the
result of LSDA is unreliable in the strong coupling regime due to its
neglect of electron correlation.

The most important thing found in this paper is the vital role of the spin
Berry phase in this geometrically frustrated system. We find the various
phases in the phases diagram are characterized(and distinguished) by their
spin Berry phase, rather than any Landau-like order parameter related to
broken symmetry. We find the spin Berry phase is responsible for the
qualitative difference in the low energy excitation spectrum of the various
states in the phase diagram. We argue that the spin Berry phase in this
system may provide the first explicit example for the concept of quantum
order and the transition between state with different spin Berry phase may
serve as the first realistic example for phase transition between states
with different quantum orders.

Another interesting thing found in this paper is the existence of an exotic
state with nonzero spin chirality but no spin ordering in a large
temperature and doping range. We find this exotic state, which break the
time reversal symmetry, support nonzero staggered current loop in the bulk
of the system. Since this state break a discreet symmetry, its phase
boundary with other phases are well defined phase transition in the Landau
sense. We propose to study this exotic state in the hole doped Na$_{1-x}$TiO$%
_{2}$ system.

As\bigskip\ a by-product of this study, we also achieve an improvement over
earlier Schwinger Boson mean field theory on the triangular antiferromagnet.
The canting ansatz proposed in this paper lead to two branches of gapless
spinon modes. We find the small gap of the third spinon mode is caused by
quantum fluctuation and our theory can reproduce the LSWT result in the
semiclassical limit.

Finally, we find our general phase diagram is consistent with experimental
result on Na$_{x}$CoO$_{2}$ system. We find singlet pairing picture for its
superconductivity may be more relevant in the small doping regime.

\bigskip {\LARGE Acknowledgement}

\bigskip The authors would like to thank members of the high Tc group at
CASTU for discussion and Z.Y.Weng for drawing our attention to the Na$_{x}$%
CoO$_{2}$ material. Works of T. Li is supported by NSFC Grant No. 90303009.

\bigskip

{\LARGE Appendix: The spinon and holon Hamiltonian} \bigskip

The spinon Hamiltonian in matrix form is

\begin{eqnarray*}
\mathbf{M}_{b} &=&\left( 
\begin{array}{cc}
\mathbf{m}_{1} & \mathbf{t}^{\ast } \\ 
\mathbf{t} & \mathbf{m}_{2}%
\end{array}%
\right) \\
&&
\end{eqnarray*}%
where

\begin{equation*}
\mathbf{m}_{1}=\left( 
\begin{array}{ccc}
\mu _{b} & x & x^{\ast } \\ 
x^{\ast } & \mu _{b} & x \\ 
x & x^{\ast } & \mu _{b}%
\end{array}%
\right) ,\mathbf{m}_{2}=\left( 
\begin{array}{ccc}
\mu _{b} & y & y^{\ast } \\ 
y^{\ast } & \mu _{b} & y \\ 
y & y^{\ast } & \mu _{b}%
\end{array}%
\right) ,\mathbf{t}=\left( 
\begin{array}{ccc}
0 & z_{_{2}} & -z_{_{1}}^{\ast } \\ 
-z_{_{2}}^{\ast } & 0 & z_{_{0}} \\ 
z_{_{1}} & -z_{_{0}}^{\ast } & 0%
\end{array}%
\right)
\end{equation*}%
in which $x=(-tF+\frac{J}{4}Q)\gamma _{k}$, $y=(-tF^{\ast }+\frac{J}{4}%
Q^{\ast })\gamma _{k}$, $z_{n}=\frac{J}{4}D\gamma _{k}e^{i\frac{2n\pi }{3}}$%
, $\gamma _{k}=\dsum\limits_{\delta }e^{ik\delta }$. The holon Hamiltonian is

\begin{equation*}
\mathbf{M}_{h}=\left( 
\begin{array}{ccc}
\mu _{_{F}} & w & w^{\ast } \\ 
w^{\ast } & \mu _{_{F}} & w \\ 
w & w^{\ast } & \mu _{_{F}}%
\end{array}%
\right)
\end{equation*}

in which $w=tQ\gamma _{\substack{ k\text{.}  \\  \\  \\ }}$

\begin{center}
\newpage

FIGURES
\end{center}

FIG. 1. The coplanar spin order at half filling and the canting state
proposed in this work.

\medskip

FIG. 2. Temperature-doping phase diagram for $t/J=1.5$. 

\medskip FIG. 3. Temperature-doping phase diagram for $t/J=-1.5$.

\medskip 

FIG. 4. Doping dependence of the net magnetization(a and c) and the
sublattice magnetization(b and d).

\bigskip 

FIG. 5. Doping dependence of the zone center gap in the coplanar state.

\medskip 

FIG. 6. Spinon dispersion of the triangular antiferromagnet as predicted in
this work(a) and the spin wave dispersion predicted by the linear spin wave
theory(b)\cite{10}. 

\bigskip 

FIG. 7. Spinon dispersion in the canted state. Note the origin of the
Brillouin zone has been moved to the gapless point. 


\bigskip

\begin{thebibliography}{99}
\bibitem{1} K. Takada, \textit{et al.}, Nature, \textbf{422} 53(2003).

\bibitem{2} G. Baskaran, Phys. Rev. Lett \textbf{91}, 097003(2003).

\bibitem{3} A. Tanaka \textit{et al.}, cond-mat/0304409.

\bibitem{4} Q. H. Wang \textit{et al.}, cond-mat/0304377.

\bibitem{5} T. Waki \textit{et al.}, cond-mat/0306036.

\bibitem{6} Y. Kobayashi \textit{et al.}, cond-mat/0306264.

\bibitem{7} T. Fujimoto \textit{et al.}, cond-mat/0307127.

\bibitem{8} T. Motohashi \textit{et al.}, Phys. Rev. B \textbf{67}, 064406
(2003).

\bibitem{9} D. J. Singh, Phy.Rev B \textbf{68}, 020503 (2003).

\bibitem{10} Th. Jolicoeur et al., Phys. Rev. B \textbf{40}, 2727(1989).

\bibitem{11} D. Yoshioka, J. Phys. Soc. Jpn. \textbf{58}, 32 (1989).

\bibitem{12} A. Mattsson, Phys. Rev. B \textbf{51},11574(1995).

\bibitem{13} Y. C. Chen, Mod. Phys. Lett. B \textbf{8}, 1253(1994).

\bibitem{14} X. G. Wen, Phys. Rev. B \textbf{65}, 165113 (2002).

\bibitem{15} C. L. Kane \textit{et al.}, Phys. Rev. B \textbf{41},
2653(1990).

\bibitem{16} P. A. Lee \textit{et al.}, Phys. Rev. B \textbf{46}, 5621
(1992).

\bibitem{17} M. Z. Hasan \textit{et al.}, cond-mat/0308438.
\end{thebibliography}
\end{document}